\documentclass[12pt]{article}
\usepackage[utf8]{inputenc}
\usepackage[utf8]{inputenc} % for pdfLaTeX
\usepackage[T1]{fontenc}
\DeclareUnicodeCharacter{202F}{\,} % thin space
\usepackage{amsmath,graphicx}
\usepackage[a4paper,margin=1in]{geometry}
\title{Controlled Growth of Bronze Telluride for Scalable Thermoelectric Energy Harvesting}

\author{
	Karthik R$^{1}$, Abhijith MB$^{2}$\thanks{Equal contribution}, Juan Gomez Quispe$^{2}$\footnotemark[1], \\
	Varinder Pal$^1$, Manas Paliwal$^1$, Ajit K Roy$^3$, Pedro Alves Da Silva Autreto$^4$, \\
	Sreeram Punathil Raman$^{1,5}$\thanks{Corresponding author: sreerampunam@metel.iitkgp.ac.in}, \\
	Pulickel M. Ajayan$^5$\thanks{Corresponding author: ajayan@rice.edu}, Chandra Sekhar Tiwary$^1$\thanks{Corresponding author: chandra.tiwary@metal.iitkgp.ac.in}
}

\date{}

\begin{document}
	
	\maketitle
	
	\begin{center}
		$^1$ Department of Metallurgical and Materials Engineering, IIT Kharagpur, India \\
		$^2$ Materials Science Center, IIT Kharagpur, India \\
		$^3$ Air Force Research Laboratory, Dayton, OH, USA \\
		$^4$ Federal University of ABC (UFABC), Santo André, Brazil \\
		$^5$ Department of Materials Science and NanoEngineering, Rice University, USA
	\end{center}
	
	\begin{abstract}
		With the growing demand for sustainable and decentralized energy solutions, thermoelectric energy harvesting has emerged as a promising technology for directly converting waste heat into electricity through solid-state, environmentally friendly means. Among copper chalcogenides, Cu$_2$Te is a notable p-type material due to its degenerate semiconducting nature and low thermal conductivity. In this study, we present a sustainable synthesis strategy for Sn-doped Cu$_2$Te referred to as bronze telluride (BT) via a chemical vapor deposition (CVD)-assisted tellurization process using pre-alloyed Cu–Sn (bronze) powder. The resulting BT exhibited an enhanced thermoelectric figure of merit (ZT) of 1 at 500 K. To assess practical applicability, BT was integrated with n-type galena (PbS) in a cascaded p–n thermoelectric module, which generated 2.8 mV across a temperature gradient of 35 K demonstrating its potential for medium- to high-temperature waste heat recovery. Furthermore, thermodynamic calculations and density functional theory (DFT) simulations provided insights into the formation mechanism of Cu$_2$Te and the thermoelectric behaviour of BT. This work introduces an efficient, scalable, and environmentally responsible pathway for developing copper-based thermoelectric materials using industrially relevant precursors.
	\end{abstract}
	
	
	\section{Introduction}
With the rising demand for sustainable and decentralized energy solutions, thermoelectric energy harvesting has gained significant attention as a clean and solid-state technology to convert waste heat into electricity[1]. This approach enables direct conversion of temperature gradients into electricity using thermoelectric materials, making it attractive for recovering energy from sources like industrial waste heat (furnaces, engines, exhausts) and domestic heaters or appliances. The efficiency of a thermoelectric material is given by its dimensionless figure of merit, $ZT = S^2\sigma T/\kappa$, where S is the Seebeck coefficient, $\sigma$ is electrical conductivity, $\kappa$ is thermal conductivity (electronic and lattice), and T is the absolute temperature. High S and $\sigma$ with low $\kappa$ are needed for high ZT and better conversion efficiency. However, these parameters are interdependent, so improving electrical conductivity can also raise thermal conductivity, making optimization challenging. [2]

Recent advancements in thermoelectric materials have demonstrated that strategies such as nano-structuring [3][4], selective doping[5][6], multiphase integration[7], and the design of hierarchical architectures[8] can significantly enhance performance. A wide range of material systems, including bismuth tellurides[9][10], tin chalcogenides [11], and half-Heusler alloys [12][13], have been extensively investigated to achieve high thermoelectric conversion efficiencies across different temperature ranges. Among the various candidates, copper-based chalcogenides with the general formula Cu$_2$X (where X = S, Se, Te) have emerged as promising thermoelectric materials due to their intrinsic degenerate semiconducting nature and earth abundance[14][15][16]. This characteristic, coupled with their capacity to suppress thermal conductivity via enhanced phonon scattering, makes them highly suitable for energy conversion across a broad temperature range [17]. Copper sulfides [18] and selenides[19] have been the subject of intensive research and have shown notable thermoelectric efficiencies, with high ZT values recorded from room temperature to elevated thermal regimes. However, copper tellurides (Cu$_2$Te) remain less explored despite their favourable base properties [20][21]. One major limitation of Cu$_2$Te is its inherent structural instability, as it undergoes several phase transitions between ambient conditions and $\approx$900 K [22]. These transitions lead to fluctuations in thermal transport, particularly by lowering thermal diffusivity at specific temperature intervals. Additionally, Cu$_2$Te is known for its highly mobile Cu$^+$ ions, which tend to vacate lattice sites easily, forming off-stoichiometric compositions (Cu$_{2-x}$Te, with 0 < x < 0.08) under equilibrium conditions from room temperature up to about 673 K [23]. This tendency introduces further complexity in maintaining consistent thermoelectric performance. Addressing this issue requires strategies that can reduce the material’s excessive carrier concentration while stabilizing its structure. Doping with tin (Sn) offers a compelling solution. Sn atoms, unlike the commonly used Ag dopants, provide an extra valence electron compared to Cu$^+$, enabling more effective charge compensation and stronger covalent bonding within the lattice. Furthermore, Sn features a lone pair of electrons in its 5s orbital, which may enrich the density of electronic states near the valence band edge [21]. This modification can result in improved Seebeck coefficients and enhanced power factors, key indicators of thermoelectric performance.

Recognizing the industrial relevance of copper and tin, key constituents of bronze and readily recoverable from recycled industrial and electronic waste, we have developed a sustainable and scalable strategy to synthesize Sn-doped Cu$_2$Te, hereafter termed bronze telluride (BT). For the first time, BT was synthesized via a chemical vapor deposition (CVD)-assisted tellurization process employing a pre-alloyed Cu–Sn powder as the precursor. In this approach, the Cu–Sn alloy powder is directly exposed to tellurium vapor in a controlled atmosphere, enabling in-situ formation of the Cu$_2$Te phase while simultaneously incorporating Sn uniformly into the crystal lattice. Compared to conventional ex-situ doping, which often leads to limited solubility and non-uniform dopant distribution, or to nanostructuring and multiphase approaches that require multiple processing steps to engineer interfaces and grain boundaries, this direct tellurization method provides superior compositional homogeneity and defect engineering in a single step. Moreover, the process is inherently adaptable for large-scale synthesis of bulk thermoelectric materials with enhanced phase purity and structural integrity, offering a viable pathway for industrial deployment. Furthermore, it aligns with environmentally responsible practices by utilizing earth-abundant and recyclable elements. The introduction of Sn serves to address key challenges associated with pristine Cu$_2$Te, such as high intrinsic carrier concentration and structural instability. Sn doping enhances the Seebeck coefficient and overall thermoelectric performance by optimizing carrier concentration and modifying the electronic structure. To evaluate its real-world applicability, the synthesized p-type BT was integrated with n-type galena (PbS) in a cascaded p–n thermoelectric module. This configuration enabled effective carrier separation and voltage generation across a temperature gradient, demonstrating the potential of the BT–PbS system for medium- to high-temperature waste heat recovery. In addition, we carried out density functional theory (DFT) calculations to clarify how Sn doping alters the electronic structure and enhances the thermoelectric properties of Cu$_2$Te by tuning its band structure and carrier concentration. We also performed thermodynamic simulations to study the phase evolution of Cu$_2$Te from bronze, providing deeper insights into its formation and stability. We measured the Seebeck coefficient, electrical and thermal conductivities across various temperatures to calculate ZT and validate the theoretical predictions, confirming that Sn addition effectively improves thermoelectric performance. Together, these theoretical, thermodynamic, and experimental results demonstrate a promising route for developing efficient, earth-abundant thermoelectric materials.

	\section{Experimental Methods}
	\subsection{CVD Growth of Bronze Telluride}
	Bronze telluride was synthesized using a chemical vapor deposition (CVD) process carried out in a ThermoFisher Lindberg Blue single-zone furnace equipped with a 1-inch quartz tube. High-purity tellurium powder (99.9\%, Sigma-Aldrich) was placed in an alumina boat upstream within the heating zone, while bronze powders were positioned in separate alumina boats at the centre of the furnace. The system was purged and maintained with argon as the carrier gas at a constant flow rate of 50 sccm throughout the growth process. The furnace temperature was ramped to 750 $^0$C at a rate of 25 $^0$C/min and held at that temperature for 10 minutes. Hydrogen gas was introduced into the chamber at 400 $^0$C and continued until the end of the growth process to assist in the tellurization reaction. After growth, the furnace was allowed to cool naturally and was opened at 400 $^0$C to facilitate faster cooling.
	
	\subsection{Material Characterization}
	
	The morphology of the as-grown bronze telluride flakes was first observed using an optical microscope (Zeiss Optical Microscope). Detailed surface morphology was examined using a field-emission scanning electron microscope (FESEM, Gemini 500, Zeiss). The elemental composition was determined through energy-dispersive X-ray spectroscopy (EDS) using an EDAX detector attached to the FESEM. Raman spectroscopy was performed using a WiTec Alpha 300R spectrometer with a 532 nm excitation laser and a 2400 g/mm grating to probe vibrational modes. Structural and crystallographic analyses were conducted using a PANalytical Empyrean X-ray diffractometer with Cu K$\alpha$ radiation, scanned over a 2$\theta$ range from 10$^0$ to 90$^0$ in gonio mode. Chemical state and binding energy information were obtained using X-ray photoelectron spectroscopy (XPS) in a ULVAC PHI-5000 VersaProbe III system. High-resolution transmission electron microscopy (HRTEM) was performed using an FEI Titan F30 operating at 300 kV to further analyze the crystal structure and lattice fringes.
	
	\subsection{Computational Details}
	The electronic properties of pristine Cu$_2$Te and its Sn-doped configurations were investigated through first-principles calculations based on Density Functional Theory (DFT), as implemented in the SIESTA package [37][38]. SIESTA is an open-source, linear-scaling code that employs a basis set of numerical atomic orbitals to represent the electronic wavefunctions, along with norm-conserving Troullier–Martins pseudopotentials in the Kleinman–Bylander form. Exchange–correlation effects were treated within the generalized gradient approximation (GGA), using the Perdew–Burke–Ernzerhof (PBE) functional [39][40]. Electron–ion interactions were described by a double-zeta polarized (DZP) basis set. A strict convergence criterion was imposed, with the density matrix tolerance set to 1×10$^{-4}$ and the maximum change in the Hamiltonian matrix limited to 1×10$^{-3}$ eV. Structural optimizations were performed using the conjugate gradient algorithm until the residual forces on all atoms were reduced below 0.01 eV/\AA. A Monkhorst–Pack [41] k-point mesh of 5×5×5 was employed for geometry optimizations, whereas a denser 10×10×10 grid was used for the calculation of the electronic band structure and projected density of states (PDOS).
	
	The electronic transport properties were assessed by solving the semiclassical Boltzmann Transport Equation (BTE) under the constant relaxation time approximation (CRTA) and the rigid band approximation (RBA), as implemented in the BoltZTraP2 code [42][43]. The BTE was applied within the linear response regime, assuming a constant scattering time ($\tau$) for all electronic states. To achieve accurate interpolation of the band structure and reliable estimation of group velocities, a dense k-point mesh of 21×21×21 was employed. This refined sampling ensures a better resolution of the energy gradients ($\Delta_k$ $\epsilon_{nk}$), which are critical for calculating the transport distribution function and related transport tensors. The computed transport quantities include the electrical conductivity (up to a prefactor of $\tau$), the Seebeck coefficient, and the electronic contribution to the thermal conductivity, all evaluated as functions of temperature. These calculations provide valuable insights into the thermoelectric performance of pristine and Sn-doped Cu$_2$Te.
	
	\subsection{Thermodynamic calculations}
	The thermodynamic calculations were performed based on the Calphad approach using FactSage software [44]. To develop the thermodynamic database for the calculations in the present system, the liquid phase was modelled using the modified quasichemical model (MQM)[45]. Gibbs energy of the solid solutions in the binaries was described using the Compound energy formalism [46]. The optimised model parameters are provided in the supplementary Table S1. In addition, Kohler-Toop interpolation was used for interpolating the binaries (Cu-Te, Sn-Te, and Cu-Sn) to the ternary system. Detailed information regarding the thermodynamic optimisation is provided in the supplementary file.

	\section{Results and Discussion}
\begin{center}
		\begin{figure}[h]
			\centering
			\includegraphics[scale=0.85]{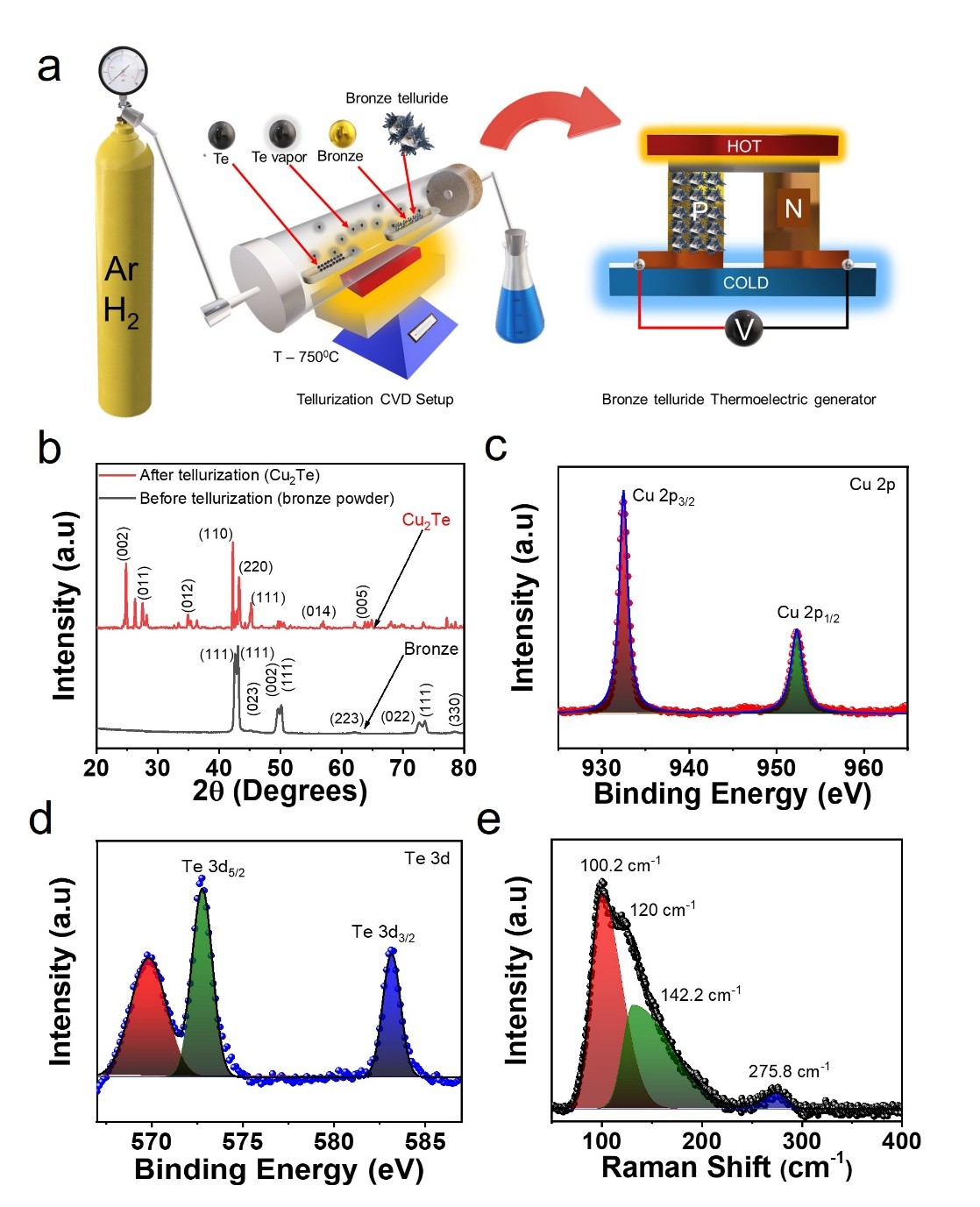}
			\caption{(a) Schematic representation of the CVD-assisted tellurization process for the synthesis of bronze telluride (Sn-doped Cu$_2$Te) for thermoelectric energy harvesting applications, (b) XRD patterns of bronze powder and bronze telluride after tellurization, (c) High-resolution XPS spectrum of Cu 2p, (d) High-resolution XPS spectrum of Te 3d, and (e) Raman spectrum of bronze telluride.}
			\label{fig:1}
		\end{figure}
	\end{center}

	Figure \ref{fig:1}a presents a schematic representation of the CVD-based synthesis route for bronze telluride, marking a novel and efficient approach for producing high-performance thermoelectric materials. To analyze the structural, oxidation, and vibrational states of bronze telluride, we conducted X-ray diffraction (XRD), X-ray photoelectron (XPS), and Raman spectroscopic studies. The XRD pattern of bronze alloy and bronze telluride formed after tellurization is presented in Figure \ref{fig:1}b. The diffraction pattern of bronze powder matches well with cubic Cu$_{3.64}$Sn$_{0.28}$ (ICSD: 629271) and hexagonal Cu10Sn3 (ICSD:103105). After tellurization, the diffraction pattern corresponds to hexagonal (P6/mmm) Cu$_2$Te (ICSD: 77055) with no detectable impurity phases such as SnTe. When compared to the standard lattice constant of pure Cu$_2$Te, which is ($\approx$8.12 \AA), a decrease of 0.943 \AA was observed in the tellurized sample. This reduction could be attributed to Sn substitution in the Cu$_2$Te lattice or strain-induced lattice contraction due to Sn incorporation without direct substitution [24]. The details regarding the calculation of lattice contestant are provided in the supporting information.  To further investigate the presence of Sn in the structure, XPS analysis was performed. The XPS survey scan (Supporting information Figure S1) confirms the presence of Cu and Te without detectable Sn peaks, indicating a very low concentration of Sn in the tellurized sample. High resolution Cu 2p XPS (Figure\ref{fig:1}c) spectra show well-resolved peaks at Cu2p$_{3/2}$ (932.4 eV) and Cu2p$^{1/2}$ (952.4 eV), corresponding to Cu$^+$ oxidation state in Cu$_2$Te. In the Te 3d XPS spectra (Figure\ref{fig:1}d), peaks at 572.7 eV (Te 3d$_{5/2}$) and 583.1 eV (Te 3d$_{3/2}$) correspond to the Te$^{2-}$ state in Cu$_2$Te. Notably, a broad peak at 570 eV is observed in the Te 3d region, which may be associated with defect-induced Te states, suggesting strain-related electronic state modification in Te [25]. The absence of any other peaks indicates the stability of Cu$_2$Te obtained through tellurization. To evaluate the phonon modes of Cu$_2$Te, we performed Raman spectroscopic studies. Figure\ref{fig:1}e presents Raman spectra of Cu$_2$Te with distinct peaks at 100.2 cm$^{-1}$, 120 cm$^{-1}$, 142.2 cm$^{-1}$ and 275.8 cm$^{-1}$. These peaks correspond to Cu-Te phonon mode (A$_1$ or E mode), Te-Te stretching mode (related to Te-rich environments), Cu$_2$Te lattice vibration (possible mixed Cu-Te modes), and Te A$_{1g}$ (characteristic of tellurium-based compounds) [26]. These results confirm the successful formation of Cu$_2$Te through tellurization of bronze with possible strain effects induced by Sn incorporation.

\begin{center}
		\begin{figure}[h]
			\centering
			\includegraphics[scale=.8]{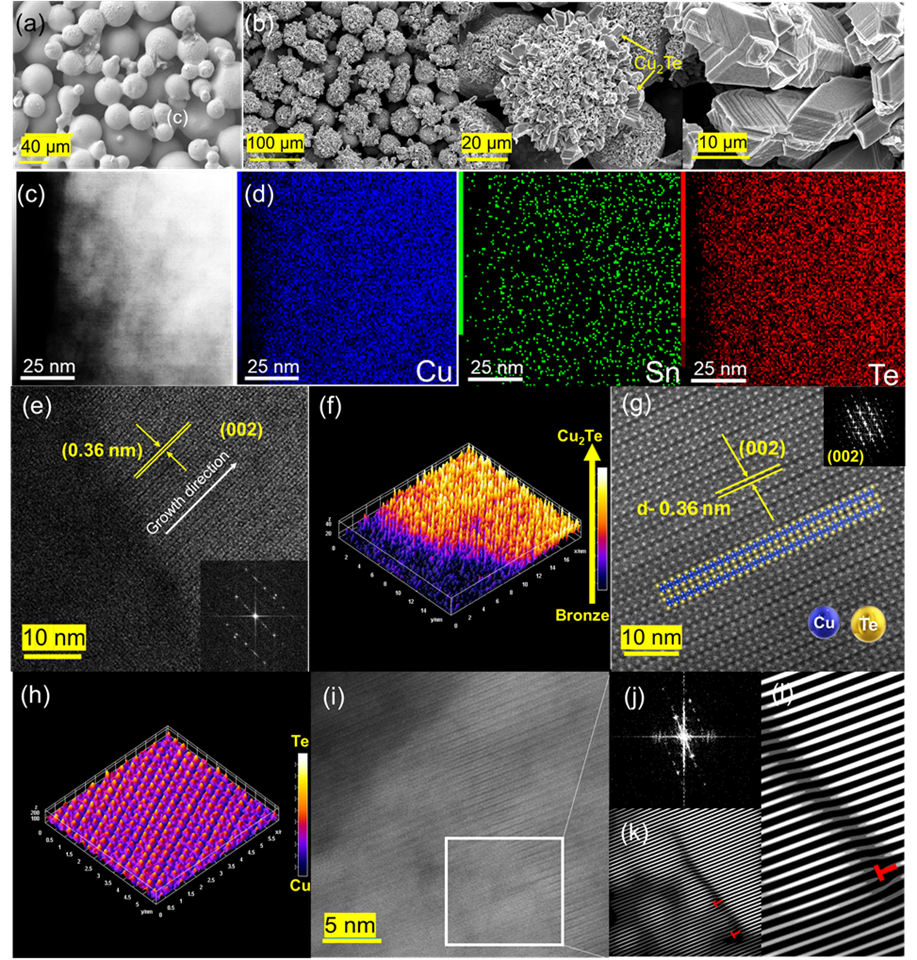}
			\caption{(a) SEM image of bronze powder particles, (b) SEM image of tellurized bronze powder showing the evolution of Cu$_2$Te with a layered morphology, (c) HRTEM image of a selected region in bronze telluride for EDS mapping, (d) EDS elemental distribution of Cu, Te, and Sn, (e) HRTEM image of bronze telluride highlighting the (002) lattice planes, (f) 3D surface plot of Figure 2e illustrating the growth of Cu$_2$Te from bronze, (g) HRTEM image of Cu$_2$Te with superimposed Cu and Te atoms for clarity, along with an FFT inset showing the (002) plane, (h) 3D surface plot of Figure 2g, (i) HRTEM image of a Cu$_2$Te region exhibiting defects, (j) FFT pattern of the defective region, and (k-l) inverse FFT (IFFT) pattern of the defective region revealing edge dislocations.}
			\label{fig:2}
		\end{figure}
	\end{center}

	To determine the morphological and structural characteristics of bronze telluride, we performed Scanning electron microscopy (SEM), Energy dispersive spectroscopy (EDS), Transmission electron microscopy (TEM) studies as shown in Figure\ref{fig:2}. The SEM image of bronze powder before tellurization (\ref{fig:2}a) reveals spherical bronze particles. After tellurization, the growth of Cu$_2$Te crystals is observed (Figure\ref{fig:2}b), displaying a layered structure with step-like morphology. While XRD and XPS studies did not detect Sn in the tellurized bronze sample, likely due to its very low concentration, high-resolution TEM (HRTEM-EDS) analysis, along with elemental mapping, was performed to assess its presence. Figure\ref{fig:2}c marks the region of interest for EDS mapping, while Figure\ref{fig:2}d shows the elemental distribution of Cu, Te, and Sn. The TEM-EDS spectra (supporting information Figure S2) reveal an Sn atomic concentration of 0.09\%, which, although minimal, may still induce lattice distortions, as observed in the XRD analysis.  To determine the growth direction of Cu$_2$Te in the tellurized sample, high-resolution transmission electron microscopy (HRTEM) studies were performed as shown in Figure\ref{fig:2}e. Fast Fourier transform (FFT) pattern analysis confirmed a lattice spacing of 0.36 nm, corresponding to the (002) plane of Cu$_2$Te. A 3D surface plot (Figure\ref{fig:2}f) further illustrates the growth of C$_2$Te from bronze. Atomic scale imaging along the (002) plane is shown in Figure\ref{fig:2}g, where Cu and Te atoms have been superimposed onto the HRTEM image for clarity. A corresponding 3D surface plot is given in Figure\ref{fig:2}h. In certain regions, discontinuities in atomic arrangements were observed as shown in \ref{fig:2}i. Further analysis using FFT (Figure\ref{fig:2}j) and Inverse (IFFT) (Figure\ref{fig:2}k and Figure\ref{fig:2}l) revealed edge dislocations in the lattice, which can be attributed to strain effects induced by Sn incorporation. These findings suggest that confirms that Sn does not substitute Cu/Te atoms, but instead creates localized lattice distortions.

	\subsection{Thermodynamic calculations}
	From TEM studies, we observed the growth of Cu$_2$Te from bronze. To understand this transformation, we conducted thermodynamic simulations to examine the formation of Cu$_2$Te from bronze powder through tellurization. The calculated phase diagram (Supplementary Figure S3) shows the phase equilibria for varying Te partial pressures and Cu content in the Cu–Sn system at 750 $^0$C. As shown, Cu$_2$Te crystallization is predicted at lower Te partial pressures and Cu mole fractions above 0.8. Further reduction of Te vapor pressure leads to the formation of BCC, D0$_3$, or FCC phases, depending on the alloy composition. These results indicate that the Te partial pressure and alloy composition in our synthesis fall within the predicted Cu$_2$Te formation region (Figure S3). Additional details on the thermodynamic calculations are provided in the Supporting Information.
	
	\subsection{Thermoelectric studies of bronze telluride}
\begin{center}
		\begin{figure}[h]
			\centering
			\includegraphics[scale=.7]{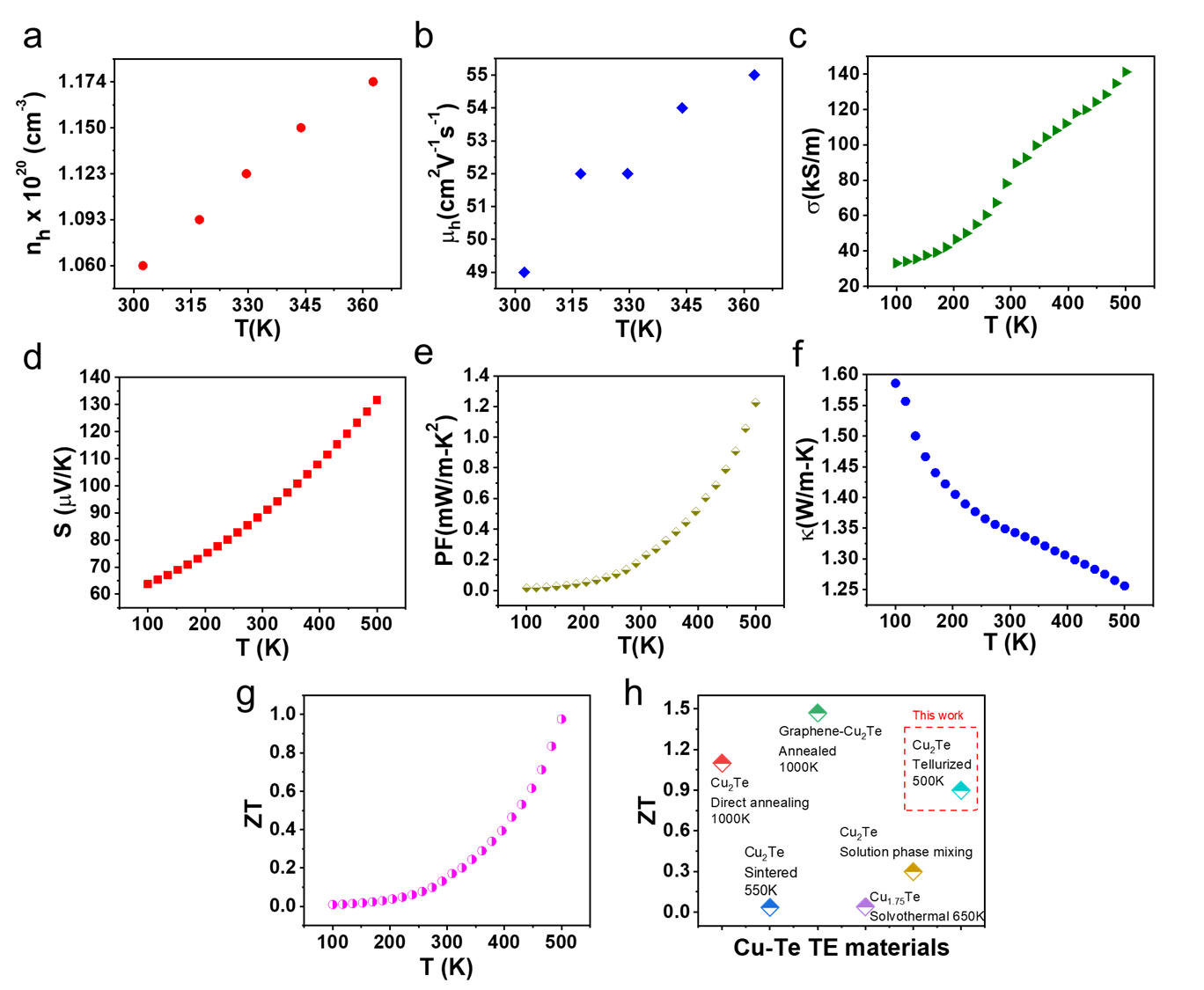}
			\caption{(a, b) Temperature-dependent carrier concentration and carrier mobility of bronze telluride, both exhibiting an increasing trend with temperature. Temperature-dependent thermoelectric properties of bronze telluride: (c) Electrical conductivity, (d) Seebeck coefficient, (e) Power factor, (f) Total thermal conductivity, and (g) Thermoelectric figure of merit (ZT). (h) Comparative analysis of ZT values for various copper telluride-based materials synthesized using different methods.}
			\label{fig:3}
		\end{figure}
	\end{center}

	To investigate the thermoelectric properties of bronze telluride, we carried out detailed measurements of carrier concentration, electrical conductivity, thermal conductivity, and the Seebeck coefficient, as illustrated in Figure\ref{fig:3}. Figure\ref{fig:3}a shows the variation of hole concentration with temperature, while Figure\ref{fig:3}b presents the corresponding hole mobility for the synthesized p-type bronze telluride. The carrier concentration exhibits a clear increase in the temperature range from 302 K to 363 K. This trend indicates thermally activated behaviour, likely due to increased ionization of shallow acceptor states or enhanced thermal excitation across a narrow bandgap[27][28]. Simultaneously, the hole mobility also shows a slight but consistent increase with temperature, rising from approximately 49 cm$^2$V$^{-1}$s$^{-1}$ at 302 K to 55 cm$^2$V$^{-1}$s$^{-1}$ at 363 K. This positive temperature dependence suggests that phonon scattering is not the dominant scattering mechanism within this temperature range; instead, grain boundary or impurity scattering effects may decrease with increasing thermal energy, allowing for improved carrier transport. The electrical conductivity (\ref{fig:3}c) exhibits an increasing trend with temperature, indicative of semiconducting behaviour. This improvement in charge transport may be linked to Sn-induced lattice distortions, which influence grain boundary scattering and contribute to better carrier mobility by reducing carrier localization. At the same time, the rise in the Seebeck coefficient (Figure\ref{fig:3}d) suggests modifications in the electronic density of states near the Fermi level. The introduction of lattice strain due to Sn presence could lead to an enhancement in the effective mass of charge carriers. Additionally, these distortions may introduce localized electronic states, further strengthening the Seebeck effect [29]. As a result, the power factor (Figure\ref{fig:3}e), which depends on both electrical conductivity and the Seebeck coefficient, also increases with temperature, reaching its peak at elevated temperatures. The thermal conductivity, shown in Figure\ref{fig:3}f, decreases with rising temperature, which can be attributed to intensified phonon scattering at higher temperatures, thereby limiting heat transport. This decline in thermal conductivity is favourable for thermoelectric applications as it helps enhance the figure of merit (ZT).  The thermoelectric figure of merit (ZT), depicted in Figure \ref{fig:3}g, approaches a value of 1 at 500 K, reinforcing the suitability of bronze telluride for low-temperature energy conversion applications. A comparative analysis (Figure \ref{fig:3}h) with other Cu$_2$Te-based thermoelectric materials synthesized through different methods indicates that the Sn-modified copper telluride, obtained via tellurization, demonstrates superior performance for low-temperature thermoelectric applications. The enhancement in thermoelectric efficiency is primarily due to the suppression of lattice thermal conductivity while retaining high electrical conductivity, thereby optimizing the balance between electronic and phononic transport properties. Although Sn does not replace Cu or Te within the lattice, its influence on local lattice distortions plays a key role in modifying the thermoelectric performance of Cu$_2$Te by increasing phonon scattering and improving charge carrier transport. These structural modifications make Sn-modified Cu$_2$Te a promising material for thermoelectric applications within the low-to-mid temperature range.
	
	\subsection{Bronze Telluride-Galena TE device}
\begin{center}
		\begin{figure}[h]
			\centering
			\includegraphics[scale=.85]{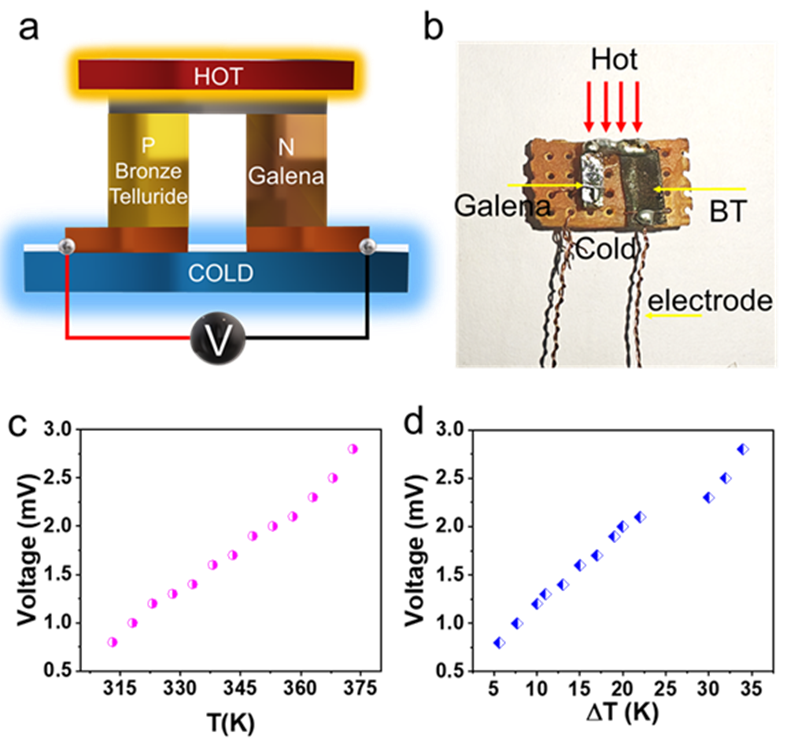}
			\caption{(a)Schematic representation of the BT-Galena thermoelectric device, illustrating the device architecture and material interfaces, (b) Digital photograph of the fabricated BT-Galena device, (c) Output voltage as a function of temperature, demonstrating the device's thermoelectric response, (d) Output voltage as a function of applied thermal gradient, confirming the thermoelectric behavior under varying thermal differentials.}
			\label{fig:4}
		\end{figure}
	\end{center}

	To demonstrate the in operando thermoelectric performance of p-type bronze telluride, we constructed a simple thermoelectric module by pairing it with n-type lead sulfide (PbS) to form a thermoelectric couple, as PbS is a well-known, earth-abundant n-type semiconductor with a high Seebeck coefficient and good compatibility in terms of processing and operating temperature with the bronze telluride. The thermoelectric properties of the n-type PbS used in this module have been previously reported in our earlier work [4]. Figure \ref{fig:4}a illustrates a schematic representation of the thermoelectric device configuration, highlighting the electrical and thermal contacts. A corresponding digital photograph of the fabricated module is shown in Figure \ref{fig:4}b, clearly depicting the physical arrangement of the p–n couple. The device was subjected to a temperature gradient across its terminals, and the thermoelectric response was monitored. As shown in Figure \ref{fig:4}c, the output voltage increases steadily with temperature, confirming the effective thermoelectric conversion behaviour of the device. This positive correlation between temperature and voltage further substantiates the functional performance of the bronze telluride as a p-type thermoelectric material. Figure \ref{fig:4}d presents the variation of the output voltage as a function of the applied temperature gradient ($\Delta$T) across the device. At a temperature difference ($\Delta$T) of 33 K, the device generated a maximum open-circuit voltage of 3.0 mV, corresponding to an effective Seebeck coefficient of approximately 90.9 $\mu$V/K. This value is notably lower than the expected combined Seebeck coefficient for the thermoelectric couple, which typically lies in the range of 290–410 $\mu$V/K, based on Sn-doped Cu$_2$Te (70–140 $\mu$V/K) and n-type PbS (–220 to –270 $\mu$V/K). Despite the reduced output, the observed voltage confirms the thermoelectric activity of the heterojunction. The lower effective Seebeck coefficient is likely due to factors such as interfacial contact resistance, material inhomogeneity, or partial electrical shorting. Therefore, power calculations were not performed, as they would introduce uncertainties. These limitations highlight the need for further optimization through interface engineering and precise control of doping and material processing. An optimized version, with precisely cut thermoelectric legs and proper device fabrication, is planned for future studies. Overall, this result demonstrates the potential of the bronze telluride–PbS module for thermoelectric power generation under practical thermal conditions.
	
	\subsection{Theoretical studies}
\begin{center}
		\begin{figure}[h]
			\centering
			\includegraphics[scale=.65]{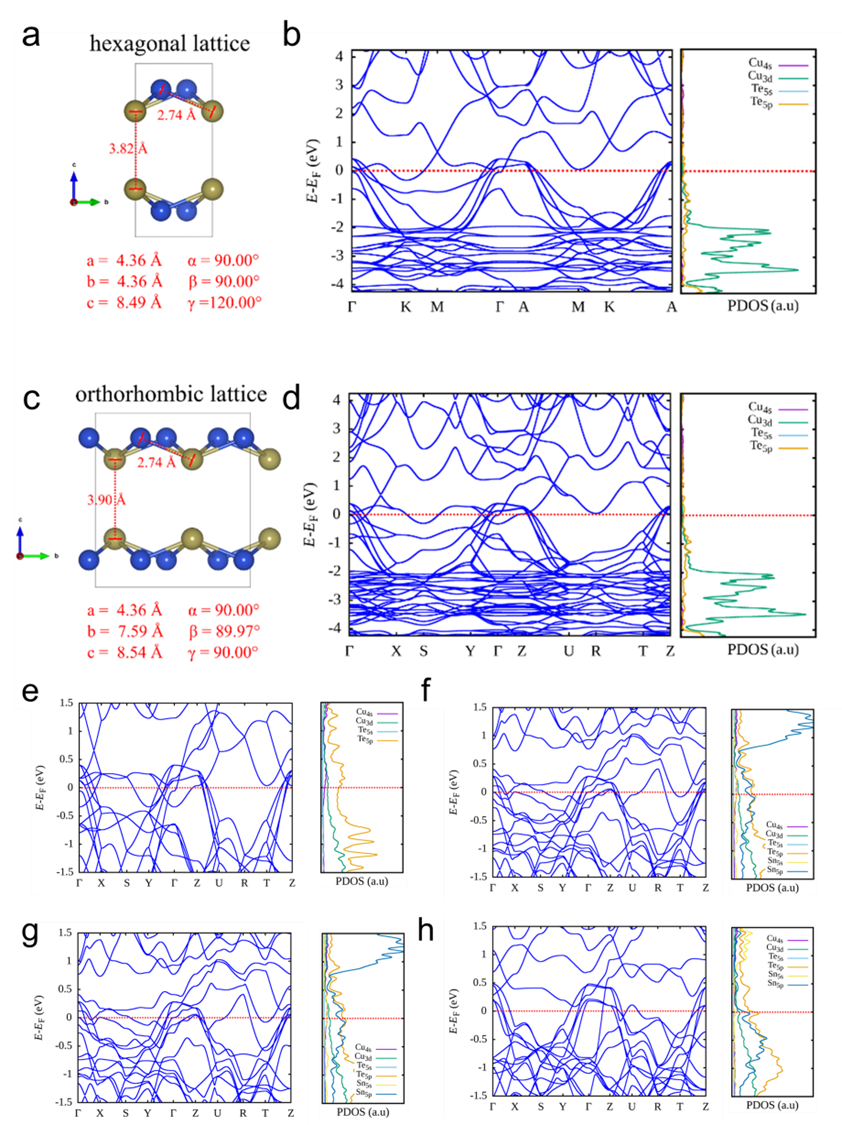}
			\caption{(a, c) Crystal structures of Cu$_2$Te in the hexagonal and orthorhombic phases, showing typical bond distances and lattice parameters, (b, d) Electronic band structure and projected density of states (PDOS) for hexagonal and orthorhombic Cu$_2$Te, respectively. Electronic band structures and projected density of states (PDOS) for the four Cu$_2$Te-based configurations analyzed in this study. (e) Pristine Cu$_2$Te (m1), (f) Sn atom incorporated in the upper interlayer region (m2), (g) Sn atom incorporated in the symmetric interlayer position (m3), (h) Sn atom embedded in the bottom interlayer region (m4). The Fermi level (E$_f$) is set to 0 eV and indicated by the dashed red line. In the PDOS panels, the atomic orbital contributions from Cu (4s, 3d), Te (5s, 5p), and Sn (5s, 5p) are shown, highlighting the changes in the electronic structure due to Sn incorporation.}
			\label{fig:5}
		\end{figure}
	\end{center}

	To investigate the impact of Sn doping on the thermoelectric properties of Cu$_2$Te, we performed density functional theory (DFT) calculations. Figures \ref{fig:5} (a-d) present the optimized atomic structures, electronic band, and projected density of states (PDOS) for pristine Cu$_2$Te in the primitive hexagonal and orthorhombic unit cells. For the hexagonal phase, the optimized lattice parameters are a = b = 4.36 \AA, c = 8.49 \AA, and $\gamma$ = 120$^0$ (Figure \ref{fig:5}a), which are in good agreement with previous theoretical and experimental studies [30–32]. The Cu-Te bond lengths of 3.82 \AA and 2.74 \AA reflect the layered nature of the Cu$_2$Te structure. In contrast, the orthorhombic phase exhibits lattice parameters of a = 4.36 \AA, b = 7.59 \AA, and c = 8.54 \AA, with all angles approximately 90$^0$ ($\alpha$ = $\beta$ = $\gamma$ = 90$^0$). This structural distortion indicates an anisotropic expansion, particularly along the b-axis, accompanied by slight variations in Cu–Te bond lengths (3.90 \AA and 2.74 \AA), suggesting a rearrangement of the atomic layers to accommodate the orthorhombic symmetry. The electronic band structures of the hexagonal and orthorhombic phases (Figures \ref{fig:5}b and \ref{fig:5}d, respectively) confirm the metallic nature of the pristine Cu$_2$Te hexagonal lattice, as evidenced by multiple bands crossing the Fermi level (E$_f$). In the orthorhombic cell, the band structure retains the key features of the hexagonal phase, indicating that the change in symmetry does not alter the intrinsic metallic behavior. Furthermore, no significant increase in band dispersion complexity is observed near E$_f$, suggesting minimal redistribution of electronic states upon the symmetry transition. The PDOS plots (Figures \ref{fig:5}b and \ref{fig:5}d) show that the states near Ef are primarily derived from Cu 3d and Te 5p orbitals. A comparison between both phases reveals a strong consistency in orbital contributions, further supporting the preservation of the electronic structure across the symmetry change. The slight redistribution in electronic density indicates that the transition from hexagonal to orthorhombic symmetry has a negligible effect on the electronic transport properties. Overall, these results demonstrate that the orthorhombic phase of Cu$_2$Te maintains the metallic nature of the material, with only minor alterations in the band structure and density of states, which are unlikely to significantly impact its thermoelectric performance compared to the hexagonal phase. The formation energies (E$_{form}$) for the different Sn-doped Cu$_2$Te structural models are presented in the supporting information. The effect of Sn atom incorporation on the electronic properties of pristine Cu$_2$Te is presented next, which shows the electronic band structures and projected density of states for the four structural models considered in this study. Figures 5e display the band structure and PDOS of pristine Cu$_2$Te, confirming its metallic character and the dominant contribution from Cu 3d orbitals. Figures 5f and 5g show the results for the models m2 y m3. In both cases, the metallic nature is preserved, and a notable contribution from Sn 5p states is observed in the conduction band, indicating that Sn could act as a donor, increasing the number of electron carriers. Figures 5h illustrate the electronic band structure and PDOS for the model m4, which is thermodynamically more stable than the pristine phase. This configuration also retains the metallic character, with an enhanced presence of Sn 5p states in the valence band, which could influence the effective mass of charge carriers and, consequently, the electronic transport properties [33].

\begin{center}
		\begin{figure}[ht]
			\centering
			\includegraphics[scale=.65]{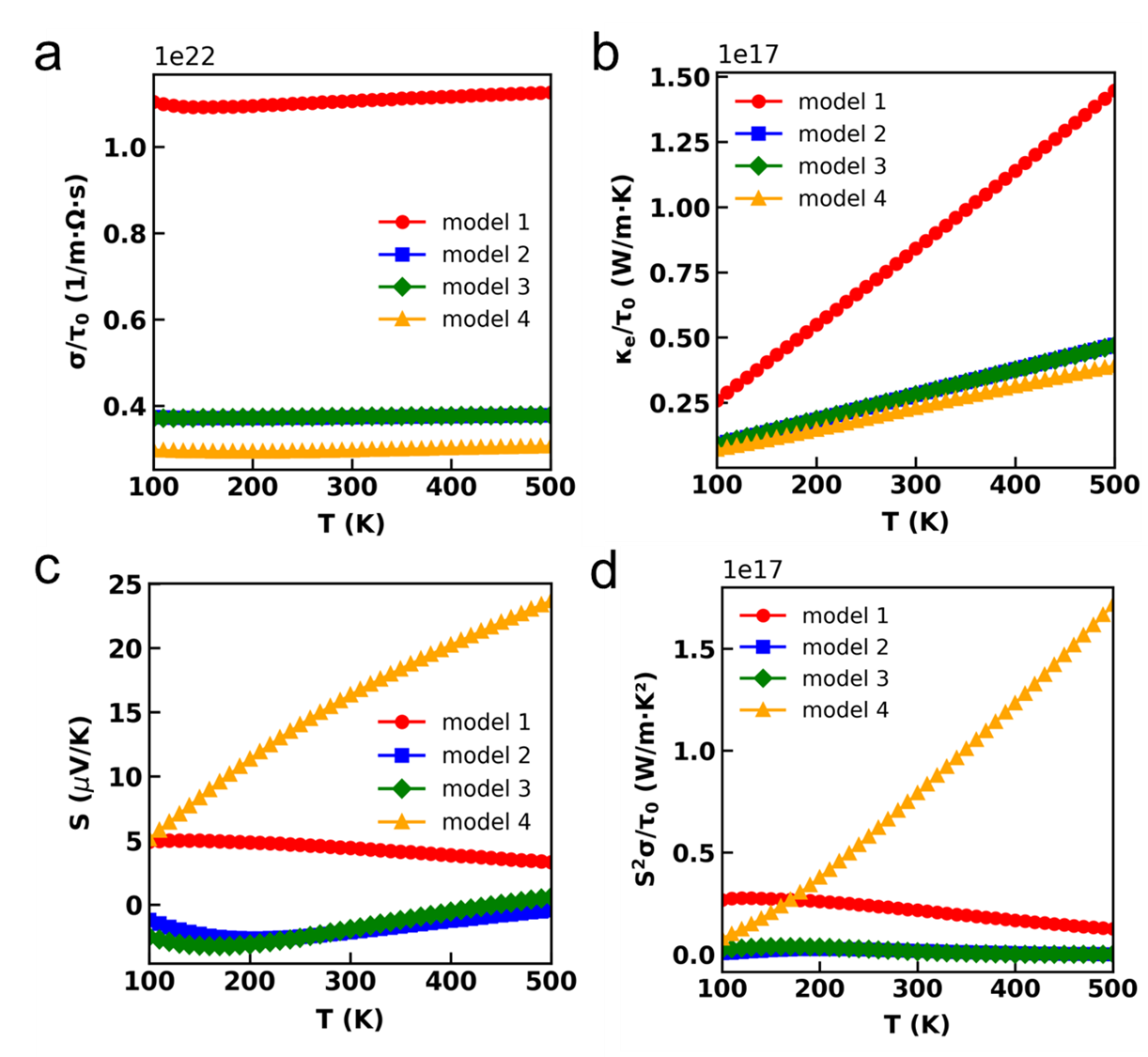}
			\caption{Temperature dependence of the thermoelectric transport properties for the four Cu$_2$Te-based systems: (a) electrical conductivity ($\sigma/\tau_0$), (b) electronic thermal conductivity ($\kappa_e/\tau_0$), and (c) Seebeck coefficient (S), (d) power factor ($S^2\sigma/\tau_0$). Pristine Cu$_2$Te (m1), Sn atom incorporated in the upper interlayer region (m2), Sn atom incorporated in the symmetric interlayer position (m3), Sn atom embedded in the bottom interlayer region (m4)}
			\label{fig:6}
		\end{figure}
	\end{center}

	Figure \ref{fig:6} presents the thermoelectric transport properties of pristine Cu$_2$Te (m1) and the three Sn-doped configurations (m2, m3, and m4), focusing on the temperature dependence of the key coefficients: electrical conductivity ($\sigma$), electronic thermal conductivity ($\kappa_e$), Seebeck coefficient (S), and power factor (PF). As shown in Figure 6a, the model m1 exhibits high electrical conductivity across the entire temperature range considered, which is consistent with its metallic nature, increasing with temperature, as confirmed by the electronic band structure (Figure 5e) [34]. Upon incorporation of a Sn atom into the interlayer region, as illustrated in models m2 and m3, a slight decrease in electrical conductivity is observed, particularly at lower temperatures. This behavior can be attributed to increased electron scattering caused by local structural distortions introduced by the Sn dopant [34]. Nevertheless, the metallic character is preserved in both configurations, as evidenced by the persistence of multiple bands crossing the E$_f$.In the case of the last model, m4, where the Sn atom is located in the lower interlayer region, the reduction in $\sigma$ is more pronounced. This decrease is associated with greater structural deformation, including significant alterations in Cu-Cu and Cu-Te bonding, as well as a shift of the Fermi level toward the valence band (Figure \ref{fig:5}d), which modifies the electronic states available for charge transport. The electronic thermal conductivity ($\kappa_e$) follows a trend like that of the electrical conductivity across all models (Figure \ref{fig:6}b), by the Wiedemann-Franz law [34]. Pristine Cu$_2$Te exhibits the highest $\kappa_e$ values, which decrease upon Sn incorporation in models m2 and m3. The same reduction is observed in model m4, where the downward shift of the Fermi level and the localization of electronic states around the Sn sites, as shown in the PDOS (see Figure \ref{fig:5}h), contribute to enhanced electron scattering and reduced electronic thermal transport capability. The Seebeck coefficient, on the other hand, exhibits a strong dependence on Sn incorporation (Figure \ref{fig:6}c). In structural models m2 and m3, where Sn 5p orbitals introduce additional states near the conduction band edge (Figures \ref{fig:5}f and \ref{fig:5}g), a slight upward shift of the E$_f$ is observed. This shift enhances the asymmetry in the density of states around the Fermi level, which could influence the thermoelectric response [35]. In contrast, model m4 shows the largest increase in the absolute values of S, which is associated with a downward shift of the E$_f$ toward the valence band, where a higher concentration of Sn 5p states is present (Figure \ref{fig:5}h). This redistribution of electronic states promotes a higher Seebeck coefficient. The power factor serves as a key metric for evaluating the overall thermoelectric performance of a material [36]. Pristine Cu$_2$Te exhibits a moderate PF (Figure \ref{fig:6}d), which decreases in models m2 and m3, primarily due to the reduction in the Seebeck coefficient, in line with the decrease in electrical conductivity. In contrast, model m4 demonstrates significantly higher Seebeck values up to two orders of magnitude greater despite its lower electrical conductivity. As a result, it achieves a moderately high PF. This result highlights a trade-off between enhancing the electronic structure and mitigating structural distortions, which may guide future strategies for targeted doping and thermoelectric optimization. The redistribution of the density of states induced by Sn incorporation plays a central role in modulating the thermoelectric properties. In models m2 and m3, the emergence of Sn 5p states near the conduction band minimum leads to an upward shift of the Fermi level, favoring n-type transport and enhanced electrical conductivity. In contrast, model m4 exhibits a significant accumulation of Sn 5p states in the valence band, which results in a downward shift of the Fermi level. This shift increases the hole concentration and enhances the Seebeck coefficient, although at the expense of lower electrical conductivity. Overall, these findings underscore the importance of carefully balancing carrier concentration and band structure engineering to maximize thermoelectric performance in Sn-doped systems.
	
	\section{Conclusion}
In summary, we have demonstrated a sustainable and scalable approach to synthesize Sn-doped copper telluride, termed bronze telluride (BT), via a CVD-assisted tellurization of pre-alloyed bronze powders. Structural and spectroscopic analyses confirmed the successful formation of a stable Cu$_2$Te phase with subtle lattice distortions induced by Sn incorporation, which effectively modulate the phonon and electronic transport properties. The thermoelectric measurements revealed that BT achieves a promising figure of merit (ZT $\approx$ 1 at 500 K) due to the combined effects of reduced thermal conductivity and enhanced Seebeck coefficient. Theoretical thermodynamic and DFT calculations provided further insights into the phase stability and the role of Sn atoms in tuning the electronic band structure and carrier concentration. Integrating BT with n-type galena (PbS) in a cascaded p–n thermoelectric module demonstrated its practical feasibility for waste heat harvesting, producing measurable output under realistic thermal gradients. Altogether, this work highlights bronze telluride as a promising candidate for low- to mid-temperature thermoelectric applications and presents an efficient route to valorize earth-abundant, recyclable precursors for sustainable energy conversion technologies. Such technologies include waste heat recovery systems in industrial processes, power generation modules for automotive exhausts, and portable or off-grid thermoelectric generators, contributing to improved energy efficiency and reduced environmental impact.
	
%	\section*{Acknowledgements}
%	[Optional section if needed]

\end{document}